\documentclass[aps,prl,preprintnumbers,amsmath,amssymb,twocolumn,superscriptaddress,showpacs,floatfix]{revtex4}

\usepackage{mathrsfs}
\usepackage{graphicx}
\usepackage{dcolumn}
\usepackage{bm}

\begin{document}

\title{Singularities, boundary conditions and  gauge link in the light cone gauge}

\author{Jian-Hua Gao}
\email{gaojh79@ustc.edu.cn}
 \affiliation{Shandong Provincial Key
Laboratory of Optical Astronomy and Solar-Terrestrial Environment,
School of Space Science and Physics, Shandong University at Weihai, Weihai 264209, China}

 \affiliation{Key Laboratory of Quark $\&$ Lepton
Physics (Central China Normal University) , Ministry of Education,
China}
\date{\today}

\vspace{-1.5in}
\vspace{1.4in}

\begin{abstract}
In this work, we first review the issues on  the singularities and the boundary conditions in light cone gauge and
how to regularize them properly. Then we will further review how these singularities and the boundary conditions
can result in the gauge link at the infinity in the light cone direction in the Drell-Yan process. Except for reviewing, we also have verified that
 the gauge link at the light cone infinity  has no dependence on the path not only for the  Abelian field but also for non-Abelian gauge field.
\end{abstract}


\maketitle


\section{introduction}
The light cone gauge was widely used as an approach to remove the redundant freedom in quantum gauge theories. The Yang-Mills theories
 were studied on the quantization  in light cone gauge by several authors \cite{Bassetto:1984dq,Leibbrandt:1983pj}. In perturbative
QCD, the collinear factorization theorems of hard processes can be proved  more  conveniently and simply in light cone gauge than in other gauges
\cite{Amati:1978by,Ellis:1978ty,Efremov:1978cu,Libby:1978qf,Libby:1978bx}. Actually
only in light cone gauge, the parton distribution functions defined in QCD hold the probability interpretation in the naive parton model\cite{Fey71}.
However in light cone gauge, when we calculate the Feynman diagrams with the gauge propagator in the perturbative theory,
we have to deal with the light cone  singularity $1/q^+ $,
\begin{eqnarray}
D_{\mu\nu}(q)=\frac{1}{q^2+i\epsilon}\left(g_{\mu\nu}-\frac{n_\mu q_\nu + n_\nu q_\mu}{q^+}\right).
\end{eqnarray}
There have been a variety of  prescriptions  suggested to handle such singularities
{\cite{Bassetto:1983bz,Lepage:1980fj,Leibbrandt:1987qv,Slavnov:1987yh,Kovchegov:1997pc} from a practical point of view.
Afterwards, it was clarified \cite{Bassetto:1981gp}  that the
gauge potential cannot be arbitrarily set to vanish at the infinity  in the light cone gauge, the spurious singularities
are physically related to the boundary conditions that one can impose on the potentials at the infinity. Different
 pole structures for regularization mean  different boundary conditions. It should be emphasized that the above conclusion does not restrict to
the light cone gauge, it holds for any axial gauges.

The non-trivial boundary conditions at the infinity in the light cone gauge also clarifies another  puzzle in
the transverse momentum dependent  structure functions of nucleons. In the
covariant gauge, in which the gauge potential vanishes at the
space-time infinity, the transverse-momentum parton distribution can be given by operator matrix elements
\cite{Collins:1981uw,Collins:1989bt,Collins:2003fm}
\begin{eqnarray}
\label{TMD}
q(x,\vec{k}_\perp)&=&\frac{1}{2}
\int \frac{d y^-}{2\pi}\frac{d^2 {\bf{y}}_\perp}{(2\pi)^2}{\rm e}^{- i x p^{+}y^{-}+i{\bf{k}}_\perp\cdot{\bf{y}}_\perp}\nonumber\\
&&\times\langle P |\bar\psi (y^-,{\bf{y}}_\perp) n\!\!\!\slash \mathcal{L}^\dagger[\infty,{\bf{y}}_\perp;y^-,{\bf{y}}_\perp ]\nonumber\\
& &\times\mathcal{L}[\infty,{\bf{0}}_\perp;0,{\bf{0}}_\perp ]\psi (0,{\bf{0}}_\perp)| P \rangle\, ,
\end{eqnarray}
where
\begin{equation}
\label{TMDGL}
\mathcal{L}[\infty,{\bf{y}}_\perp; y^-,{\bf{y}}_\perp]\equiv
P \exp \left(- i g \int_{y^-}^\infty d \xi^{-} A^+ ( \xi^-, {\bf{y}}_\perp)
\right) \, ,
\end{equation}
is the gauge link or wilson line to ensure the gauge invariance of the matrix elements.
Such gauge link is produced  from final state interactions between the struck quark and the target
spectators.   It has been  verified in Ref.\cite{Collins:1992kk} that  the presence of  the gauge link is essential for
the non-vanishing  Sivers function, which is the main mechanism of single spin asymmetry at low transverse momentum in high energy  collisions.
However if we naively  choose the light cone gauge $A^+ =0$ in the above definition,  it seems as if the gauge link in
Eq.~(\ref{TMDGL}) would disappear, which would result in the  final interaction or Siver's function
vanishing. It seems as if  different gauges lead to contradictory results.
Since physics should not depend on the gauge we choose, there must be  something we missed in the above. Such contradiction was
solved by Ji and Yuan in \cite{Ji:2002aa} where they found that the final state interaction effects
can be recovered properly  in the light cone
gauge by introducing a transverse gauge link at the
light cone infinity. Then in \cite{Belitsky:2002sm}, Belitsky, Ji, and
Yuan demonstrated  how  the transverse gauge link  can be produced from the transverse components
of the gauge potential at the light cone
infinity at the  leading twist level. Further in \cite{Gao:2010sd}, we  derived such transverse gauge link within a more regular and general
method. It was found that  the gauge link at light cone infinity naturally
arises from the contribution of the pinched poles:
one is from the quark propagator and the other is hidden in the gauge vector field in the light cone gauge. It is just the pinched poles that
pick out the contribution of the gauge potential at the light cone infinity.
Actually in \cite{Gao:2010sd},  a more general gauge link  over the hypersurface
at the light cone infinity was derived, which is beyond the transverse direction. Besides, there are also other relevant works on the transverse
gauge link in the literature \cite{Boer:2003cm,Idilbi:2008vm,Cherednikov:2011ku}.

In this paper, we will devote ourselves to
reviewing the above works and putting them together  with the emphasis  on  mathematical rigor.
However, through the reviewing, we will try to discuss them in a different way or  point of view,
which can be also regarded as the complementary to the previous work. Except for reviewing, we also have verified that
 the gauge link at the light cone infinity  has no dependence on the path not only for the  Abelian field but also for non-Abelian gauge field,
 which has not been  discussed  in the previous works.

We will organize the paper as follows: in Sec.~\ref{notations}, we  present
some  definitions and notations which will be used in our paper.
In Sec.~\ref{singularity}, we  discuss how the singularity can arise in the light cone gauge,
how different singularities correspond to different boundary conditions and how we regularize them properly.
In Sec.~\ref{derivation},  we  derive the transverse gauge link or more general gauge link in light cone gauge in Drell-Yan process.
In Sec.~\ref{path}, we verify that  the gauge link at the light cone infinity  has no dependence on the path for non-Abelian gauge field.
In Sec.~\ref{summary},  we  give a brief summary.

\section{definitions and notations}
\label{notations}
In our work, we will choose  the light cone coordinate system
by  introducing two lightlike vectors $n^{\mu}$ and $\bar{n}^{\mu}$ and two transverse spacelike vectors $n_{\perp1}^{\mu}$ and ${n}_{\perp2}^{\mu}$
\begin{eqnarray}
 n^\mu &=&\frac{1}{\sqrt{2}}\left(1,0,0,1\right)\equiv[0,1,{\bf{0}}_{\perp}],\ \\
 \bar n^\mu &=&\frac{1}{\sqrt{2}}\left(1,0,0,-1\right)\equiv[1,0,{\bf{0}}_{\perp}]\\
 n_{\perp1}^{\mu}&=&(0,0,1,0)\equiv[0,0,1,0] ,\\
 n_{\perp2}^{\mu}&=&(0,0,0,1)\equiv[0,0,0,1]
\end{eqnarray}
where we have used square brackets `[ ]' to denote the components in the light-cone coordinate,
compared with the usual Cartesian coordinate denoted by the parentheses `( )'.
In such coordinate system, we can write any  vector $k^{\mu}$ as $[k^{+},k^{-},{\bf{k}}_{\perp}]$ or $[k^{+},k^{-},{k}_{\perp1},{k}_{\perp2}]$,
where $k^{+}=k\cdot n,\ k^{-}=k\cdot \bar{n},\ k_{\perp1}=n_{\perp1}\cdot k,\ k_{\perp2}=n_{\perp2}\cdot k$.

Since we will consider the non-Abelian gauge field all through our paper, we will  use the usual compact notations
 for the non-Abelian field potential and strength, respectively,
\begin{eqnarray}
  {A}_{\mu}\equiv {A}^a_{\mu}t^a,\ \ F_{\mu\nu}\equiv F^a_{\mu\nu} t^a,
\end{eqnarray}
where $t^a$ is the fundamental representation of the generators of the gauge symmetry group.

For the sake of conciseness, we would like to introduce some further notations. We will decompose  any momentum vector $k^{\mu}$ and the gauge potential vector $A^{\mu}$,
as the following
\begin{eqnarray}
k^{\mu}&=&{\tilde{k}}^{\mu}+xp^{\mu},\ \ A^{\mu}={\tilde{A}}^{\mu}+ A^{+} \bar{n}^{\mu}
\end{eqnarray}
where ${\tilde{k}}^{\mu}=[0,k^{-},{\bf{k}}_{\perp}]$, $x=k^+/p^+$, and $\tilde{A}^{\mu}=[0,A^{-},{\bf{A}}_{\perp}]$.
Meanwhile, for any coordinate vector $y^{\mu}$ , we will make the following decomposition,
\begin{eqnarray}
y^{\mu}&=&{\dot{y}}^{\mu}+y^{-} n^{\mu}
\end{eqnarray}
where $\dot{y}^{\mu}=[y^{+},0,{\bf{y}}_{\perp}]$. With these notations,
it is very easy to show  $k\cdot y=\tilde{k}\cdot\dot{y} +xp^{+}y^{-}$. In the light cone gauge $A^+=0$,  the gauge vector  $ A^{\mu}={\tilde{A}}^{\mu}$.
 When  no confusion could arise, we will write $y^\mu$ as $[y^-,\dot{y}]$  for simplicity.

\section{singularities and boundary conditions in the light cone gauge}
\label{singularity}
In this section, we will review how the singularities  appear in the light cone gauge,  how they are related to the boundary conditions
of the gauge potential at the light cone infinity and how we can regularize them in a proper  way consistent with the boundary conditions.
Although this section is mainly based on the literature
\cite{Bassetto:1981gp} and \cite{Belitsky:2002sm}, there are also a few  differences from them.
For example, we will discuss the non-Abelian gauge field from the beginning to the end,
while in the original works, only the Abelian gauge field was emphasized.
Besides we will make the maximal gauge  fixing from the point of view of
linear differential equation.

With the light cone gauge condition $n_\mu  A^\mu =0$,
let us consider the non-Abelian counterpart of Maxwell equations,
\begin{eqnarray}
\label{maxwell}
D_\mu F^{\mu\nu} =-j^\nu
\end{eqnarray}
where $ F_{\mu\nu} =\partial_\mu A_\nu -\partial_\nu A_\mu -ig \left[A_\mu, A_\nu\right],\ \ j^\nu=\bar\psi t^a\gamma^\nu\psi t^a$.
We can rewrite the above equations in another form
\begin{equation}
\label{maxwell-1}
\partial_\mu \partial^\mu A^\nu -\partial^\nu \partial_\mu  A^\mu = -J^\nu
\end{equation}
where we have defined $J^\nu\equiv j^\nu +ig  [A_\mu,F^{\mu\nu}] +ig\partial_\mu [A^\mu, A^\nu]$. Contracting both  sides of Eq.(\ref{maxwell-1}) with $n_\nu$
and taking the light cone gauge condition into account yields
\begin{equation}
n_\nu \partial^\nu  \partial_\mu \tilde A^\mu =n_\nu J^\nu.
\end{equation}
Integrating the above equation gives rise to
\begin{equation}
\label{constraint}
\tilde \partial_\mu \tilde A^\mu(+\infty, \tilde x)- \tilde\partial_\mu \tilde A^\mu(-\infty, \tilde x)
=\int_{-\infty}^{+\infty}dx^- n_\nu J^\nu
\end{equation}
Since $\int_{-\infty}^{+\infty}dx^- n_\nu J^\nu=0$, in general, need not be true, we can not arbitrarily choose
both $A^\mu(+\infty, \tilde x)=0$ and $A^\mu(-\infty, \tilde x)=0$ at the same time. One of these boundary conditions
can be arbitrarily chosen while the other one must be subjected to satisfy the constraint (\ref{constraint}). This
is just why we can not choose the boundary conditions arbitrarily in light cone gauge. In fact, this conclusion holds for any axial gauges.
From the Fourier transforms of Eq.(\ref{maxwell-1}),
\begin{equation}
k^2 \mathcal{A}^\nu -k^\nu (k\cdot \mathcal{A})=-\mathcal{J}
\end{equation}
and together with the light cone gauge condition, it is easy to  obtain the formal solutions
\begin{equation}
\label{solution}
\tilde A^\mu=\int d^4k \frac{e^{ik\cdot x}}{k^2}\left(-\mathcal{\tilde J}^\mu+\frac{\tilde k^\mu}{k^+}{\mathcal{J}}^+\right)
\end{equation}
where $\mathcal{A}^\mu $ and $\mathcal{J}^\mu$ are the Fourier transforms of $A^\mu$ and $J^\mu$,respectively.
It is obvious that there is an extra singularity at $k^+=0$ in the solution (\ref{solution}). If we assume that the currents
are regular at $k=0$, it is easy to verify that
the different pole prescriptions correspond to different boundary conditions. In our paper, we will
consider three different boundary conditions,
\begin{eqnarray}
\label{poles}
\textrm{Advanced} &:& \ \tilde{A}(+\infty,\dot{y})=0 \nonumber\\
\textrm{Retarded} &:&\ \tilde{A}(-\infty,\dot{y})=0 \nonumber\\
\textrm{Antisymmetric} &:& \ \tilde{A}(-\infty,\dot{y})+\tilde{A}(\infty,\dot{y})=0
\end{eqnarray}
which correspond to three different pole structures, respectively,
\begin{eqnarray}
\label{regularization}
\ \frac{1}{k^{+}-i\epsilon},\ \ \ \
\ \frac{1}{k^{+}+i\epsilon},\ \ \ \ \frac{1}{2}\left(\frac{1}{k^{+}+i\epsilon}+\frac{1}{k^{+}-i\epsilon}\right).
\end{eqnarray}
where the last prescription is just the conventional principal value regularization.
In the next section,  we will deal with the
 Fourier transform of the gauge potential,
\begin{eqnarray}
\tilde{\cal{A}}_\mu(k^+,\dot{y})\equiv\int_{-\infty}^{\infty}dy^{-}\textrm{e}^{ik^{+}y^{-}}\tilde{A}_{\mu}(y^{-},\dot{y})
\end{eqnarray}
In order to pick out the contribution of the gauge potential at the infinity,
we need a mathematical trick by manipulating this integration by parts,
\begin{eqnarray}
\label{integration}
& &\int_{-\infty}^{\infty}dy^{-}\textrm{e}^{ik^{+}y^{-}}\tilde{A}_{\mu}(y^{-},\dot{y})\nonumber\\
&=&\frac{i}{k^{+}}\int_{-\infty}^{\infty}dy^{-}\textrm{e}^{ik^{+}y^{-}}\partial^{+}\tilde{A}_{\mu}(y^{-},\dot{y})
\end{eqnarray}
where $\partial^{+}=\partial_-={\partial}/{\partial y^-}$. We will see that once we choose the prescriptions (\ref{regularization}) according to
the boundary conditions (\ref{poles}), we will obtain the gauge link at the light cone infinity.

We have seen that we cannot choose the boundary conditions arbitrarily, now we will discuss how to fix the gauge freedom as maximally as possible.
These have been also discussed in the appendix in Ref.\cite{Belitsky:2002sm}, we will take them into account from the point of view of
differential equations.
Under a general gauge transformation,  the gauge potential transforms as,
\begin{equation}
A_\mu \rightarrow S^{-1} A_\mu S +\frac{i}{g}S^{-1}\partial_\mu S
\end{equation}
In order to eliminate   the light cone component  $n\cdot A=0$, we obtain the gauge transformation by solving the equation,
\begin{equation}
n^\mu \partial_\mu S = ig n^\mu A_\mu S
\end{equation}
This equation is an  ordinary linear differential equation, whose  solution is well known,
\begin{eqnarray}
S&=& {P}\left\{\textrm{exp}\left[ig\int_{x_0^+}^{x^+}  n ^\mu A_\mu\left(\xi,x^-,x_{\perp1},x_{\perp2}\right)d\xi \right]\right\}\nonumber\\
& &\times\tilde S(x^-,x_{\perp1},x_{\perp2})
\end{eqnarray}
where  $S(x^-,x_{\perp1},x_{\perp2})$  is an arbitrary unitary matrix which does not depend on $x^+$. This freedom allows us  to
set  one of the residual three components of $A_\mu$ zero on the three dimensional hyperplane at $x^+=x_0^+$. Without loss of generality, we can set
$A^-(x_0^+,x^-,x_\perp)=0$ by solving the following equation
\begin{equation}
\bar n^\mu \partial_\mu \tilde S = ig \bar n^ \mu A_\mu \tilde S.
\end{equation}
The solution is given by
\begin{eqnarray}
\tilde S &=& {P}\left\{\textrm{exp}\left[ig\int_{x_0^-}^{x^-} d\xi  \bar n ^\mu  A_\mu\left(x_0^+,\xi,x_{\perp1}, x_{\perp2}\right)\right]\right\}\nonumber\\
& &\times S_\perp(x_{\perp1},x_{\perp2}).
\end{eqnarray}
There is still an arbitrary unitary matrix which depends only on $x_\perp$. We can use this freedom
 to further set one of the residual transverse components of the gauge potential zero, e,g, $A_{\perp1}$=0, at
the two dimensional hyperplane ($x^+=x_0^+$, $x^-=x_0^-$) by solving
\begin{equation}
n_{\perp1 }^\mu \partial_\mu S_\perp = ig n_{\perp1 }^\mu A_\mu S_{\perp}(x_{\perp1},x_{\perp2})
\end{equation}
The solution is given by
\begin{eqnarray}
S_{\perp}&=& {P}\left\{\textrm{exp}\left[ig\int_{x_{0\perp1}}^{x_{\perp1}} d\xi n_{\perp1} ^\mu
 A_\mu\left(x_0^+,x_0^-,\xi,x_{\perp2}\right)\right]\right\}\nonumber\\
& & \times S_{1\perp}(x_{\perp2}).
\end{eqnarray}
We can continue to set the only left transverse components  $A_{\perp2}=0$ at the straight line [$x^+=x_0^+$, $x^-=x_0^-$, $x_{\perp1}=x_{0\perp1}$]
by solving
\begin{equation}
n_{\perp2 }^\mu \partial_\mu S_{1\perp} = ig n_{\perp2 }^\mu A_\mu S_{1\perp}(x_{\perp2})
\end{equation}
The solution is given by
\begin{eqnarray}
\label{S1perp}
S_{1\perp}&=& {P}\left\{\textrm{exp}\left[ig\int_{x_{0\perp1}}^{x_{\perp1}} d\xi n_{\perp2} ^\mu
 A_\mu\left(x_0^+,x_0^-,x_{0\perp2},\xi\right)\right]\right\}\nonumber\\
& &\times S_{2\perp}
\end{eqnarray}
With only a trivial global gauge transformation left,  we have maximally fixed our gauge freedom.
Although we cannot choose the boundary conditions of the gauge potential arbitrarily in the light cone gauge,
the constraint that the field strengths should vanish at the infinity requires that the gauge potential must be a pure gauge,
\begin{equation}
\label{nonabelian}
{A}_{\mu}=\frac{1}{ig}\omega^{-1}{\partial}_{\mu}\omega
\end{equation}
where $\omega=\textrm{exp}(i\phi)$ with $\phi\equiv \phi^a t^a$. We can expand the above pure gauge  as
\begin{eqnarray}
\label{A-expansion}
A_\mu&=&\frac{1}{ig}\omega^{-1}{\partial}_{\mu}\omega\nonumber\\
&=&\partial_\mu\phi +\frac{i}{2!}\left[\partial_\mu \phi,\phi\right]
+\frac{i^2}{3!}\left[\left[\partial_\mu \phi,\phi\right],\phi\right]+\cdot \cdot \cdot \ \ \
\end{eqnarray}

\section{Gauge link in the light cone gauge in Drell-Yan process}
\label{derivation}

\begin{figure}
\begin{center}
\includegraphics[width=6cm]
{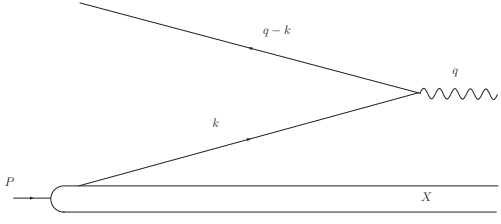} \caption{The tree diagram in Drell-Yan process} \label{DY0}
\end{center}
\end{figure}

\begin{figure}
\begin{center}
\includegraphics[width=6cm]
{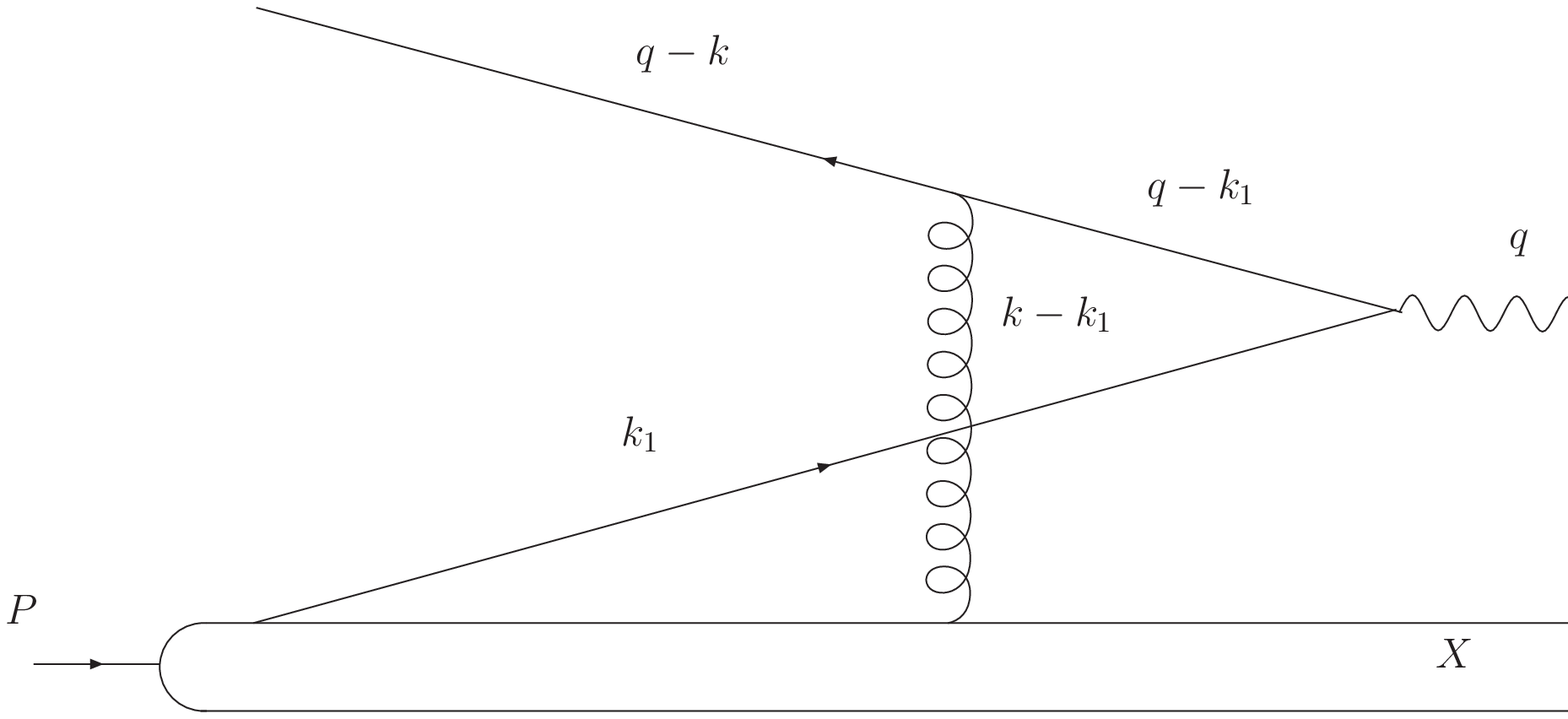} \caption{The one-gluon exchange diagram in Drell-Yan process} \label{DY1}
\end{center}
\end{figure}

\begin{figure}
\begin{center}
\includegraphics[width=6cm]
{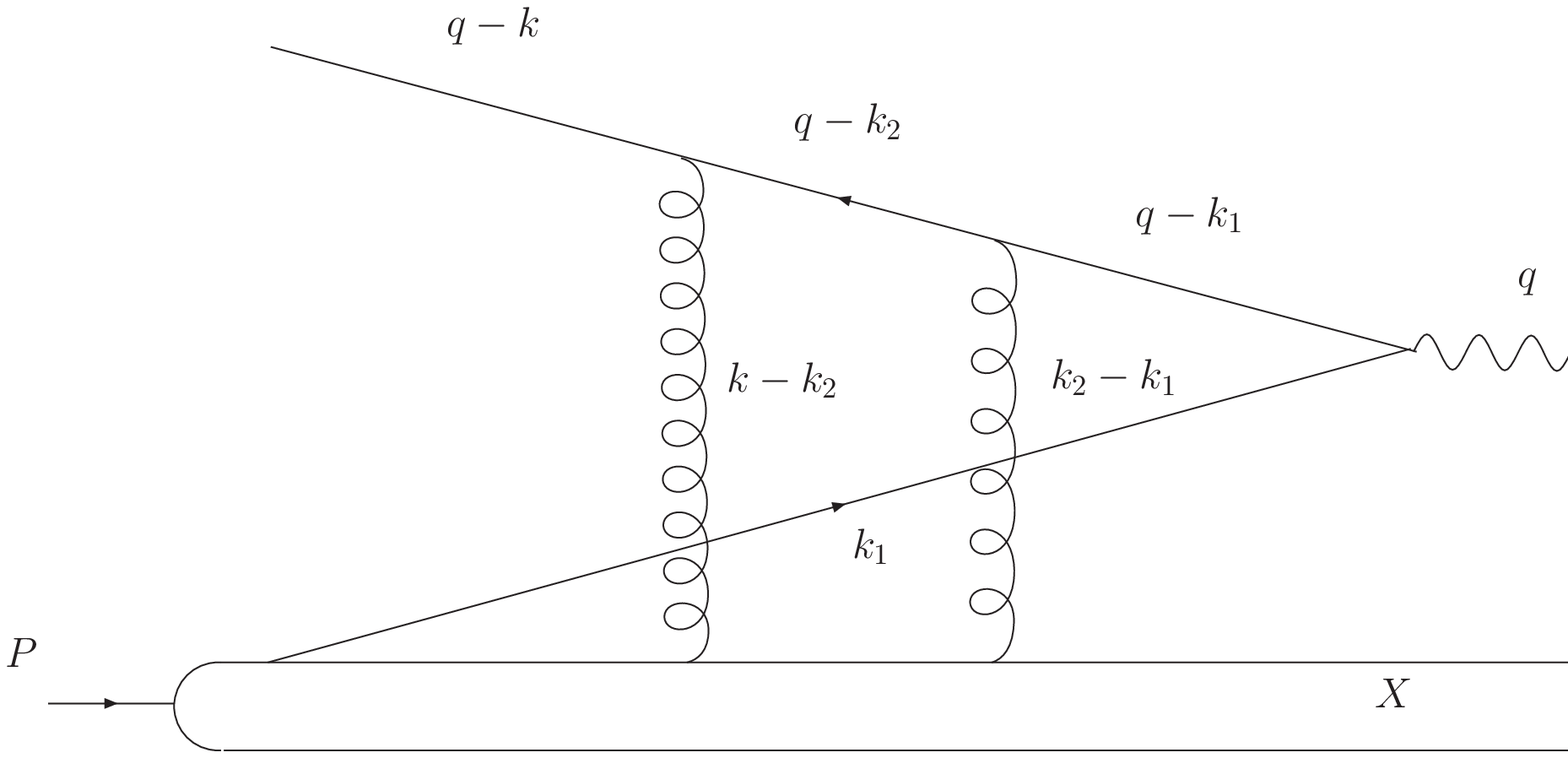} \caption{The two-gluon exchange diagram in Drell-Yan process} \label{DY2}
\end{center}
\end{figure}

\begin{figure}
\begin{center}
\includegraphics[width=6cm]
{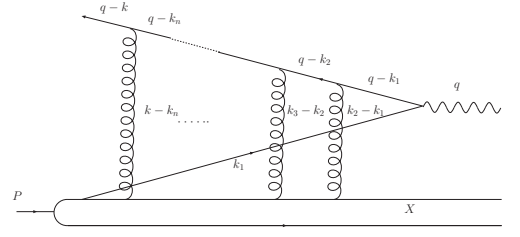} \caption{The $n$-gluon exchange diagram in Drell-Yan process} \label{DYn}
\end{center}
\end{figure}

In this section, we will review how the singularities and boundary conditions in the light cone gauge can result in the gauge link
at the light cone infinity. Since the detailed derivation of transverse gauge link had  been made for semi-inclusive deeply inelastic scattering
\cite{Belitsky:2002sm,Gao:2010sd},
we will discuss  the Drell-Yan process in details in order to avoid  total repeating. For simplicity, we will set
the target to be  a nucleon and the projectile be just an antiquark.

The tree scattering amplitude of Drell-Yan process corresponding to Fig.\ \ref{DY0} reads
\begin{equation}
M_{0}^\mu =\bar{u}(q-k)\gamma^\mu \langle X|\psi(0)|P\rangle\,,
\end{equation}
where $k$ denotes the momentum of initial quark from the proton  $P$ with the momentum $p$,
and $q-k$ and $q$ are the momenta of the anti-quark and virtual photon, respectively.

The one-gluon amplitude in the light cone gauge corresponding to Fig.\ \ref{DY1} reads,
\begin{eqnarray}
M_{1}^\mu &=&\int \frac{d^4k_{1}}{(2\pi)^4} d^4y_{1}\ e^{i(k-k_{1})\cdot y_{1}}
\bar{u}(q-k)\gamma^{\rho_{1}}\frac{{q\!\!\!\slash-k\!\!\!\slash}_{1}\!\!}{(q-k_{1})^2+i\epsilon}\nonumber\\
& &\times\langle X|\tilde{A}_{\rho_{1}}(y_{1})\gamma^{\mu}\psi(0)|P\rangle\,.
\end{eqnarray}
In order to obtain the leading twist contribution, we only need the pole contribution in  the quark propagator,
\begin{eqnarray}
\label{Mhat}
\hat{M}_{1}^\mu &=&\int \frac{d^3\tilde{k}_{1}}{(2\pi)^4} d^3\dot{y}_{1} \frac{p^{+}dx_{1}}{2\pi}dy_{1}^{-}
\ e^{i(\tilde{k}-\tilde{k}_{1})\cdot \dot{y}_{1}+i({x}-{x}_{1})p^{+}y^{-}}\nonumber\\
&&\times \bar{u}(q-\hat{k})
\gamma^{\rho_{1}}\frac{q\!\!\!\slash\!\!-{\hat{k}\!\!\!\slash}_{1}}{2p\cdot(\hat{k}_{1}-q)}\frac{1}{(x_{1}-\hat{x}_{1}-i\epsilon)}\nonumber\\
& &\times\langle X|\tilde{A}_{\rho_{1}}(y_{1})\gamma^{\mu}\psi(0)|P\rangle
\end{eqnarray}
where  $\hat{M}_{1}^\mu$ with an extra $\hat{\ }$ denotes that  only the pole contribution
is kept and $\hat{k}_1 \equiv[\hat{x}_1p^+,k_1^-,k_{1\perp}]$ with $\ \hat{x}_1=\hat{k}^{+}/p^{+}=x_{B}+k_{\perp}^2/2p\cdot (k_1-q)\ $,
which is determined by the on-shell condition $\ (q-\hat{k}_1)^2=0\ $. In Eq.(\ref{Mhat}), we have  separated the integral over $x_1$ and $y^-_1$ from the others in order to finish integrating them out
first. Now we need to choose a specific boundary condition for the gauge potential $\tilde{A}_{\rho}$ at
the infinity. Let us start with the advanced boundary condition $\tilde{A}(+\infty,\dot{y})=0 $. Using Eq.\ (\ref{integration})
for the advanced boundary condition, we  have
\begin{eqnarray}
\label{M1hat1DY-A}
\hat{M}_{1}^\mu &=&\int \frac{d^3\tilde{k}_{1}}{(2\pi)^4}\int d^3\dot{y}_{1}\ \int \frac{dx_{1}}{2\pi}\int dy_{1}^{-}\nonumber\\
& &\times e^{i(\tilde{k}-\tilde{k}_{1})\cdot \dot{y}_{1}}
e^{i({x}-{x}_{1})p^{+}y^{-}}\bar{u}(q-k)\gamma^{\rho_{1}}\nonumber\\
&&\times \frac{{q\!\!\!\slash-\hat{k}\!\!\!\slash}_{1}\!\!}{2p\cdot(\hat{k}_{1}-q)}
\frac{1}{(x_{1}-\hat{x}_{1}-i\epsilon)}\frac{i}{({x}-{x}_{1}-i\epsilon)}\nonumber\\
& &\times\langle X|\partial^{+}\tilde{A}_{\rho_{1}}(y_{1})\gamma^{\mu}\psi(0)|P\rangle\hspace{0.5cm}
\end{eqnarray}
Finish  integrating over $x_1$ and $y_1^{-}$:
\begin{eqnarray}
\label{contour5}
&&\int \frac{dx_{1}}{2\pi} dy_{1}^{-}e^{i({x}-{x}_{1})p^{+}y^{-}}
\frac{1}{(x_{1}-\hat{x}_{1}-i\epsilon)}\nonumber\\
& &\times\frac{i}{(x-{x}_{1}-i\epsilon)} \partial^{+}\tilde{A}_{\rho_{1}}(y_{1}) \nonumber\\
&=&-\int dy_{1}^{-}
\left(\theta(-y^{-})e^{i({x}-\hat{x}_{1})p^{+}y^{-}}+\theta(y^{-})\right)\nonumber\\
& &\times \frac{1}{{x}-\hat{x}_{1}} \partial^{+}\tilde{A}_{\rho_{1}}(y_{1})\nonumber\\
&=&\frac{1}{{x}-\hat{x}_{1}}\tilde{A}_{\rho_{1}}(-\infty,\dot{y}_{1})+ \textrm{higher twist}\,.
\end{eqnarray}
where only the leading term in the Tailor expansion of the phase factor $e^{i({x}-\hat{x}_{1})p^{+}y^{-}}$
is kept, because the other terms are proportional to $({x}-\hat{x}_{1})^n=\left[k_{\perp}^2/2p\cdot (k+q)-k_{1\perp}^2/2p\cdot (k_1+q)\right]^n$ $(n\ge 1)$,
which will contribute at higher twist level.

However if we choose  the retarded boundary conditions, we can have
\begin{eqnarray}
\label{M1hat1DY}
\hat{M}_{1}^\mu &=&\int \frac{d^3\tilde{k}_{1}}{(2\pi)^4}\int d^3\dot{y}_{1}\ \int \frac{dx_{1}}{2\pi}\int dy_{1}^{-}\nonumber\\
& &\times e^{i(\tilde{k}-\tilde{k}_{1})\cdot \dot{y}_{1}}
e^{i({x}-{x}_{1})p^{+}y^{-}}\bar{u}(q-k)\gamma^{\rho_{1}}\nonumber\\
& &\frac{{q\!\!\!\slash-\hat{k}\!\!\!\slash}_{1}\!\!}{2p\cdot(\hat{k}_{1}-q)}
\frac{1}{(x_{1}-\hat{x}_{1}-i\epsilon)}\frac{i}{({x}-{x}_{1}+i\epsilon)}\nonumber\\
& &\times \langle X|\partial^{+}\tilde{A}_{\rho_{1}}(y_{1})\gamma^{\mu}\psi(0)|P\rangle.
\hspace{0.5cm}
\end{eqnarray}
Integrating out $x_1$ and $y_1^{-}$ first yields
\begin{eqnarray}
\label{contour4}
&&\int \frac{dx_{1}}{2\pi}\int dy_{1}^{-}e^{i({x}-{x}_{1})p^{+}y^{-}}\frac{1}{(x_{1}-\hat{x}_{1}-i\epsilon)}\nonumber\\
& &\times\frac{i}{(x-{x}_{1}+i\epsilon)}
\partial^{+}\tilde{A}_{\rho_{1}}(y_{1}) \nonumber\\
&=&-\int dy_{1}^{-}
\left(\theta(-y^{-})e^{i({x}-\hat{x}_{1})p^{+}y^{-}}-\theta(-y^{-})\right)\nonumber\\
& &\times\frac{1}{{x}-\hat{x}_{1}}\partial^{+}\tilde{A}_{\rho_{1}}(y_{1})\nonumber\\
&=& \textrm{higher twist}\,.
\end{eqnarray}
We can see  the retarded boundary condition does not result in leading twist contribution in the Drell-Yan process.
If we choose the antisymmetric boundary condition, which corresponds to the principal value regularization, we  obtain
\begin{eqnarray}
\label{contour3}
&&\int \frac{dx_{1}}{2\pi}\int dy_{1}^{-}e^{i({x}-{x}_{1})p^{+}y^{-}}
\frac{1}{(x_{1}-\hat{x}_{1}+i\epsilon)}\nonumber\\
& &\times\textrm{PV}\frac{i}{(x-{x}_{1})}
\partial^{+}\tilde{A}_{\rho_{1}}(y_{1}) \nonumber\\
&=&\int dy_{1}^{-}
\frac{1}{2}\left(2\theta(y^{-})e^{i({x}-\hat{x}_{1})p^{+}y^{-}}-\theta(y^{-})+\theta(-y^{-})\right)\nonumber\\
& &\times\frac{1}{{x}-\hat{x}_{1}}\partial^{+}\tilde{A}_{\rho_{1}}(y_{1})\nonumber\\
&=&\frac{1}{{x}-\hat{x}_{1}}\tilde{A}_{\rho_{1}}(+\infty,\dot{y}_{1})+ \textrm{higher twist}\,,
\end{eqnarray}
where $\mbox{PV}$ denotes principal value.
In the above derivation, we notice that the presence of the pinched poles are
necessary  to pick up the gauge potential at the light cone infinity. Actually these pinched poles have selected the so-called Glauber modes of the gauge field\cite{Idilbi:2008vm}.
Although there is no leading twist contribution in the retarded boundary condition, it was  shown in \cite{Belitsky:2002sm}, that all the final state interactions have been
encoded into the initial state light cone wave functions.
In principal value regularization, the  final state scattering effects appear only through  the gauge link , while in advanced regularization,
it  appear through  both the gauge link and initial light cone wave
functions. In the following, we will only concentrate on the advanced boundary condition.
Only keep leading twist contribution and inserting Eq.\ (\ref{contour5}) into Eq.\ (\ref{M1hat1DY-A}), we  have
\begin{eqnarray}
\hat{M}_{1}&=&\int \frac{d^3\tilde{k}_{1}}{(2\pi)^4} d^3\dot{y}_{1}\
e^{i(\tilde{k}-\tilde{k}_{1})\cdot \dot{y}_{1}}
 \bar{u}(q-{k})
\gamma^{\rho_{1}}\frac{q\!\!\!\slash-{\hat{k}\!\!\!\slash}_{1}}{2p\cdot (\hat{k}_{1}-q)}\nonumber\\
& &\times\frac{1}{{x}-\hat{x}_{1}}\langle X|\tilde{A}_{\rho_{1}}(-\infty,\dot{y}_{1})\psi(0)|P\rangle\,.\hspace{0.3cm}
\end{eqnarray}
Using Eq.\ ({\ref{A-expansion}}), only keeping the first Abelian term  and performing the integration by parts over $\dot{y}_{1}$, we  obtain
\begin{eqnarray}
\hat{M}_{1}&=&\int\frac{d^3\tilde{k}_{1}}{(2\pi)^4}d^3\dot{y}_{1}\
e^{i(\tilde{k}-\tilde{k}_{1})\cdot \dot{y}_{1}}
\bar{u}(q-{k})
(\tilde{k}\!\!\!\slash-\tilde{k}\!\!\!\slash_{1})\nonumber\\
& &\times\frac{q\!\!\!\slash-{\hat{k}\!\!\!\slash}_{1}}{2p\cdot (\hat{k}_{1}-q)}
\frac{-i}{{x}-\hat{x}_{1}}\langle X|\phi(-\infty,\dot{y}_{1})\psi(0)|P\rangle.\hspace{0.7cm}
\end{eqnarray}
We can calculate these Dirac algebras and   finally obtain,
\begin{eqnarray}
\label{M1last}
\hat{M}_{1}
&=&\bar{u}(q-k)
\langle X|i\phi(-\infty,{0})\psi(0)|P\rangle\,.
\end{eqnarray}
where we have dropped all the higher twist contributions.
Now let us further consider  the two-gluon exchange  scattering amplitude plotted in Fig. \ref{DY2},
\begin{eqnarray}
M_{2}^\mu &=&\int \frac{d^4k_{2}}{(2\pi)^4}\frac{d^4k_{1}}{(2\pi)^4} d^4y_{2}d^4y_{1}
\ e^{i(k-k_{2})\cdot y_{2}+i(k_{2}-k_{1})\cdot y_{1}}\nonumber\\
&&\times \bar{u}(q-k)
\gamma^{\rho_{2}}\frac{q\!\!\!\slash-{k\!\!\!\slash}_{2}}{(q-k_{2})^2+i\epsilon}
\gamma^{\rho_{1}}\frac{q\!\!\!\slash-{k\!\!\!\slash}_{1}}{(q-k_{1})^2+i\epsilon}\nonumber\\
& &\times\langle X|\tilde{A}_{\rho_{2}}(y_{2})\tilde{A}_{\rho_{1}}(y_{1})\gamma^\mu\psi(0)|P\rangle\,.
\end{eqnarray}
Analogously to the case of  $M_{1}^\mu$, we will only keep the pole contribution,
\begin{eqnarray}
\hat{M}_{2}^\mu&=&\int \frac{d^3\tilde{k}_{2}}{(2\pi)^3}\frac{d^3\tilde{k}_{1}}{(2\pi)^3} d^3\dot{y}_{2}d^3\dot{y}_{1}
\frac{p^{+}dx_{2}}{2\pi}\frac{p^{+}dx_{1}}{2\pi} dy_{2}^{-}dy_{1}^{-}\nonumber\\
&&\times e^{i(\tilde{k}-\tilde{k}_{2})\cdot \dot{y}_{2}+i(\tilde{k}_{2}-\tilde{k}_{1})\cdot \dot{y}_{1}
+i({x}-{x}_{2})p^{+}y_{2}^{-}+i({x}_{2}-{x}_{1})p^{+}y_{1}^{-}}\nonumber\\
&&\times \bar{u}(q-k)
\gamma^{\rho_{2}}\frac{q\!\!\!\slash-{\hat{k}\!\!\!\slash}_{2}}{2p\cdot(\hat{k}_{2}-q)}
\gamma^{\rho_{1}}\frac{q\!\!\!\slash-{\hat{k}\!\!\!\slash}_{1}}{2p\cdot(\hat{k}_{1}-q)}\nonumber\\
& &\times\frac{1}{(x_{2}-\hat{x}_{2}-i\epsilon)}\frac{1}{(x_{1}-\hat{x}_{1}-i\epsilon)}\nonumber\\
&&\times \langle X|\tilde{A}_{\rho_{2}}(y_{2})\tilde{A}_{\rho_{1}}(y_{1})\gamma^\mu\psi(0)|P\rangle\,.
\end{eqnarray}
With the regularization (\ref{integration}) and (\ref{regularization}), we can  integrate
out $x_2$ and $y^{-}_2$ first,
\begin{eqnarray}
\hat{M}_{2}^\mu
&=&\int \frac{d^3\tilde{k}_{2}}{(2\pi)^3}\frac{d^3\tilde{k}_{1}}{(2\pi)^3} d^3\dot{y}_{2}d^3\dot{y}_{1}
\frac{p^{+}dx_{1}}{2\pi} dy_{1}^{-}\nonumber\\
& &\times  e^{i(\tilde{k}-\tilde{k}_{2})\cdot \dot{y}_{2}+i(\tilde{k}_{2}-\tilde{k}_{1})\cdot \dot{y}_{1}
+i({x}-{x}_{1})p^{+}y_{1}^{-}}\nonumber\\
&&\times \bar{u}(q-k)
\gamma^{\rho_{2}}\frac{q\!\!\!\slash-{\hat{k}\!\!\!\slash}_{2}}{2p\cdot(\hat{k}_{2}-q)}
\gamma^{\rho_{1}}\frac{q\!\!\!\slash-{\hat{k}\!\!\!\slash}_{1}}{2p\cdot(\hat{k}_{1}-q)}\nonumber\\
& &\times\frac{1}{(x-\hat{x}_{2}-i\epsilon)}\frac{1}{(x_{1}-\hat{x}_{1}-i\epsilon)} \nonumber\\
&&\times\langle X|\tilde{A}_{\rho_{2}}(-\infty,\dot{y}_{2})\tilde{A}_{\rho_{1}}(y_{1})\gamma^\mu\psi(0)|P\rangle\,.
\end{eqnarray}
Further integrating out $x_1$ and $y^{-}_1$ gives rise to
\begin{eqnarray}
\hat{M}_{2}^\mu
&=&\int \frac{d^3\tilde{k}_{2}}{(2\pi)^3}\frac{d^3\tilde{k}_{1}}{(2\pi)^3} d^3\dot{y}_{2}d^3\dot{y}_{1}
\  e^{i(\tilde{k}-\tilde{k}_{2})\cdot \dot{y}_{2}+i(\tilde{k}_{2}-\tilde{k}_{1})\cdot \dot{y}_{1}}\nonumber\\
&&\times \bar{u}(q-k)
\gamma^{\rho_{2}}\frac{q\!\!\!\slash-{\hat{k}\!\!\!\slash}_{2}}{2p\cdot(\hat{k}_{2}-q)}
\gamma^{\rho_{1}}\frac{q\!\!\!\slash-{\hat{k}\!\!\!\slash}_{1}}{2p\cdot(\hat{k}_{1}-q)}\nonumber\\
& &\times\frac{1}{(x-\hat{x}_{2}-i\epsilon)}\frac{1}{(x-\hat{x}_{1}-i\epsilon)} \nonumber\\
&&\times\langle X|\tilde{A}_{\rho_{2}}(-\infty,\dot{y}_{2})\tilde{A}_{\rho_{1}}(-\infty,\dot{y}_{1})\gamma^\mu\psi(0)|P\rangle
\end{eqnarray}
Using Eq.\ ({\ref{A-expansion}}), only keeping the first Abelian  term, we  have
\begin{eqnarray}
\hat{M}_{2}^\mu
&=&\int \frac{d^3\tilde{k}_{2}}{(2\pi)^3}\frac{d^3\tilde{k}_{1}}{(2\pi)^3} d^3\dot{y}_{2}d^3\dot{y}_{1}
\  e^{i(\tilde{k}-\tilde{k}_{2})\cdot \dot{y}_{2}+i(\tilde{k}_{2}-\tilde{k}_{1})\cdot \dot{y}_{1}}\nonumber\\
&&\times \bar{u}(q-k)
\gamma^{\rho_{2}}\frac{q\!\!\!\slash-{\hat{k}\!\!\!\slash}_{2}}{2p\cdot(\hat{k}_{2}-q)}
\gamma^{\rho_{1}}\frac{q\!\!\!\slash-{\hat{k}\!\!\!\slash}_{1}}{2p\cdot(\hat{k}_{1}-q)}\nonumber\\
& &\times\frac{1}{(x-\hat{x}_{2}-i\epsilon)}\frac{1}{(x-\hat{x}_{1}-i\epsilon)} \nonumber\\
&&\times\langle X|\tilde{\partial}_{\rho_{2}}\phi(-\infty,\dot{y}_{2})\tilde{\partial}_{\rho_{1}}\phi(-\infty,\dot{y}_{1})\gamma^\mu\psi(0)|P\rangle\,.
\end{eqnarray}
Using  the integration by parts, we can integrate out  $\tilde{k}_2$ and $\dot{y}_2$ and obtain

\begin{eqnarray}
\hat{M}_{2}^\mu
&=&\int \frac{d^3\tilde{k}_{1}}{(2\pi)^3} d^3\dot{y}_{1}
\  e^{i(\tilde{k}_{2}-\tilde{k}_{1})\cdot \dot{y}_{1}}\nonumber\\
& &\times \bar{u}(q-k)
\gamma^{\rho_{1}}\frac{q\!\!\!\slash-{\hat{k}\!\!\!\slash}_{1}}{2p\cdot(\hat{k}_{1}-q)}
\frac{1}{(x-\hat{x}_{1}-i\epsilon)} \nonumber\\
&&\times \langle X|\frac{i}{2}\tilde{\partial}_{\rho_{1}}\phi^2(-\infty,\dot{y}_{1})\gamma^\mu\psi(0)|P\rangle
\end{eqnarray}
Further  by  integrating over
$\tilde{k}_1$ and $\dot{y}_1$ and we finally obtain
\begin{eqnarray}
\hat{M}_{2}^\mu&=& \bar{u}(q-k)
\langle X|\frac{i^2}{2!}\phi^2(-\infty,{0})\gamma^\mu\psi(0)|P\rangle
\end{eqnarray}
It should be noted that we have neglected the higher twist contributions in the above derivation.
It is obvious that the procedure from ${M}_{1}$ to ${M}_{2}$ can be easily generalized to higher order amplitudes.
For example, the general $n$-gluon exchange
amplitude $M_{n}$ in Fig. \ref{DYn}  reads
\begin{eqnarray}
\hat{M}_{n}
&=&
\int\prod_{j=1}^{n}\frac{d^3{\tilde{k}_{j}}}{(2\pi)^3}d^3\dot{y}_{j}
e^{i(\tilde{k}_{n}-\tilde{k}_{n-1}) \cdot\dot{y}_{n}+\ ...\  +i(\tilde{k}_{2}-\tilde{k}_{1})\cdot\dot{y}_{2}}\nonumber\\
&&\times
\prod_{j=1}^{n}
\frac{p^{+}d{x}_{j}}{2\pi}dy^{-}_{j}
e^{i({x}_{n+1}-{x}_{n})p^{+} {y}_{n}^{-}+\ ...\  +i({x}_{2}-{x}_{1})p^{+}{y}_{1}^{-}}\nonumber\\
&&\times \bar{u}(q-k)
\gamma^{\rho_{n}}\frac{q\!\!\!\slash-{\hat{k}\!\!\!\slash}_{n}}{2p\cdot(\hat{k}_{n}-q)}
\ \ ...\ \gamma^{\rho_{1}}\frac{q\!\!\!\slash-{\hat{k}\!\!\!\slash}_{1}}{2p\cdot(\hat{k}_{1}-q)}\nonumber\\
& &\times\frac{1}{(x_{n}-\hat{x}_{n}-i\epsilon)}\ \ ...\ \frac{1}{(x_{1}-\hat{x}_{1}-i\epsilon)}  \nonumber\\
&&\times\langle X|\tilde{A}_{\rho_{n}}(y_{n})\ ...\ \tilde{A}_{\rho_{1}}(y_{1})\psi(0)|P\rangle\,.
\end{eqnarray}
We  first finish integrating from $x_{n},y^{-}_{n}$ to $x_{1},y^{-}_{1}$ one by one. Keeping the leading twist
contribution, we have,
\begin{eqnarray}
\hat{M}_{n}
&=&
\int\prod_{j=1}^{n}\frac{d^3{\tilde{k}_{j}}}{(2\pi)^3}d^3\dot{y}_{j}
e^{i(\tilde{k}_{n+1}-\tilde{k}_{n}) \cdot\dot{y}_{n}+\ ...\ +i(\tilde{k}_{2}-\tilde{k}_{1})\cdot\dot{y}_{2}}\nonumber\\
&&\times \bar{u}({k}+q)
\gamma^{\rho_{n}}\frac{{\hat{k}\!\!\!\slash}_{n}\!\!+q\!\!\!\slash}{2p\cdot(\hat{k}_{n}+q)}
\ \ ...\
\gamma^{\rho_{1}}\frac{{\hat{k}\!\!\!\slash}_{1}\!\!+q\!\!\!\slash}{2p\cdot(\hat{k}_{1}+q)}\nonumber\\
& &\times \frac{1}{(\hat{x}_{n+1}-\hat{x}_{n})}\ \ ...\ \
\frac{1}{(\hat{x}_{2}-\hat{x}_{1})} \nonumber\\
&&\times \langle X|\tilde{\partial}_{\rho_{n}}\phi(-\infty,\dot{y}_{n})\tilde{\partial}_{\rho_{n-1}}\phi(-\infty,\dot{y}_{n-1})\nonumber\\
& &\times \ ...\ \tilde{\partial}_{\rho_{1}}\phi(-\infty,\dot{y}_{1})\psi(0)|P\rangle\,.
\end{eqnarray}
where we have used Eq.\ ({\ref{A-expansion}}) again and only kept the first Abelian  term.
Integrating one by one over from momenta from $\tilde{k}_n$ and $\dot{y}_n$ to $\tilde{k}_1$ and $\dot{y}_1$ one by one, we  finally have
\begin{eqnarray}
\hat{M}_{n}
&=&
\bar{u}(q-{k})\langle X|\frac{i^n}{n!}\phi^n(-\infty,0)\psi(0)|P\rangle\,.
\end{eqnarray}
As  final step,  we  resum to all orders and obtain
\begin{eqnarray}
\sum_{n=0}^{\infty}\hat{M}_{n}
&=&\bar{u}(q-k)\langle X|\textrm{exp}\left(i\phi(-\infty,0)\right)\psi(0)|P\rangle\nonumber\\
&=&\bar{u}(q-k)\langle X|\omega\left(-\infty,0\right)\psi(0)|P\rangle
\end{eqnarray}
The light cone infinity  $y^-=-\infty$ instead of $y^-=+\infty$ reflects that the phase factor arises from the initial
interaction rather than from the final interaction. Now we need to express $\omega$ as the function of $A_\mu$ by solving
Eq.(\ref{nonabelian}) at the light cone infinity. We can rewrite Eq.(\ref{nonabelian}) in the partial differential form,
\begin{equation}
\label{nonabelian-1}
\tilde {\partial}_{\mu}\omega(-\infty,\dot x)=ig \tilde {A}_{\mu}(-\infty,\dot x)\omega(-\infty,\dot x)
\end{equation}
This equation cannot be solved unless certain integrability conditions are satisfied. In Sec.~\ref{path}, we will show that
$F_{\mu\nu}=0$ is  the right integrability condition, which is assumed to be always satisfied for the gauge field at the infinity.
The solution is exactly the gauge link that we want,
\begin{eqnarray}
\label{omega-link}
\omega(-\infty,\dot x)&=&P \exp \left( i g \int_{-\dot\infty}^{\dot x}d \dot\xi_\mu \tilde A^\mu ( -\infty, \dot\xi)\right)\nonumber\\
&=&\mathcal{L}\left[-\infty,\dot x;-\infty,-\dot \infty\right] \, ,
\end{eqnarray}
where we have chosen  $\omega(-\infty,-\dot \infty)=1$ which can be always achieved by using the residual global gauge transformation   $S_{2\perp}$ in (\ref{S1perp}).
It follows that
\begin{eqnarray}
\hat{M}_{n}
&=&
\bar{u}(q-{k})\langle X|\psi(0)\mathcal{L}\left[-\infty,\dot 0; -\infty,-\dot \infty\right]|P\rangle\,.
\end{eqnarray}
It should be emphasized that the gauge link we obtain here
is  over the hypersurface at the light cone infinity along any path integral, not restricted along the transverse direction.
Let us verify this independence in the next section.

\section{Path independence of the gauge link}
\label{path}
In this section, we will show that the gauge link (\ref{omega-link})  is the solution of Eq.(\ref{nonabelian-1}) with the integrability condition $F_{\mu\nu}=0$.
However we would like to  prove   a more general conclusion here. We will
verify that  the arbitrary gauge link connecting $x_0$ with $x$,
\begin{eqnarray}
\label{omega}
& &\omega(s;x_0,x)\nonumber\\
&=&{P}\left\{\textrm{exp}\left[ig\int_0^{s} ds_1\frac{dy^{\nu_1}}{ds_1}A_{\nu_1}\left(y(s_1;x_0,x)\right)\right]\right\},
\end{eqnarray}
where $s_1$ denotes the path parameter  with the constraints
\begin{eqnarray}
\label{y0s}
y(0;x_0,x)=x_0,\ \ \  y(s;x_0,x)=x.
\end{eqnarray}
is the solution of the equation
\begin{eqnarray}
\label{dif}
{\partial}_{\mu}\omega =ig{A}_{\mu}\omega
\end{eqnarray}
under the integrability condition $F_{\mu\nu}=0$.

For the sake of briefness, we introduce some compact notations,
\begin{eqnarray}
\mathscr{A}(s_1)&\equiv&\frac{dy^{\nu_1}}{ds_1}A_{\nu_1}\left(y(s_1;x_0,x)\right),\\
\mathscr{A}_\mu(s_i)&\equiv& \partial_\mu y^{\nu_i} A_{\nu_i}\left(y(s_i;x_0,x)\right),\\
\mathscr{F}_\mu(s_i)&\equiv& \frac{dy^{\nu_i}}{ds_1} \partial_\mu y^\rho F_{\rho\nu_i}\left(y(s_i;x_0,x)\right)
\end{eqnarray}
We can expand $\omega(s)$ as
\begin{eqnarray}
\label{omega-expansion}
\omega(s)= \phi_0  + ig \phi_1 + (ig)^2 \phi_2 + (ig)^3 \phi_3 + \cdot\cdot \cdot
\end{eqnarray}
where we have defined
\begin{eqnarray}
\phi_0&=&1\\
\phi_1&=&\int_0^{s} ds_1\mathscr{A}(s_1)\\
\phi_2&=&\int_0^{s} ds_1 \int_0^{s_1} ds_2 \mathscr{A}(s_1)\mathscr{A}(s_2)\\
\phi_3&=&\int_0^{s} ds_1 \int_0^{s_1} ds_2 \int_0^{s_2} ds_3 \mathscr{A}(s_1)\mathscr{A}(s_2)\mathscr{A}(s_3)
\end{eqnarray}
We have suppressed all the dependence on the $x_0$ and $x$.
In the following, we devote ourselves to calculating $\partial_\mu \omega$ in details. In order to do that, we need to
calculate the partial derivative of each term in the expansion (\ref{omega-expansion}). The partial derivative of the zeroth order is trivially zero.
Let us calculate the first order,
\begin{eqnarray}
\partial_\mu \phi_1
&=&\int_0^{s} ds_1\frac{d(\partial_\mu y^{\nu_1})}{ds_1}A_{\nu_1}\left(y(s_1)\right)\nonumber\\
& &+\int_0^{s} ds_1\frac{dy^{\nu_1}}{ds_1} \partial_\mu y^\rho\frac{ \partial}{\partial y^\rho} A_{\nu_1}\left(y(s_1)\right)
\end{eqnarray}
where it should be noted that $\partial_\mu \equiv \partial /\partial x^\mu$ here and  we must distinguish it from  $\partial /\partial y^\mu$.
Using the differential chain type rule and the definition $F_{\mu\nu} =\partial_\mu A_\nu -\partial_\nu A_\mu -ig \left[A_\mu, A_\nu\right]$, we  have
\begin{eqnarray}
\label{partial-1}
\partial_\mu \phi_1
&=&\int_0^{s} ds_1\frac{d(\partial_\mu y^{\nu_1})}{ds_1}A_{\nu_1}\left(y(s_1)\right)\nonumber\\
& &+\int_0^{s} ds_1 \partial_\mu y^\rho \frac{d}{ds_1} A_\rho\left(y(s_1)\right)\nonumber\\
& &+\int_0^{s} ds_1\frac{dy^\nu}{ds_1} \partial_\mu y^\rho\left\{F_{\rho\nu_1}\left(y(s_1)\right)\right.\nonumber\\
& &\left.\hspace{2cm}+ig \left[ A_\rho\left(y(s_1)\right), A_\nu\left(y(s_1)\right)\right]\right\}\nonumber\\
&=&\mathscr{A}_\mu(s)-\mathscr{A}_\mu(0)\nonumber\\
& &+\int_0^{s} ds_1\left\{\frac{}{} \mathscr{F}_\mu(s_1)
+ig\left[ \mathscr{A}_\mu\left(s_1\right), \mathscr{A}\left(s_1\right)\right]\right\}\hspace{1cm}
\end{eqnarray}
Now let us turn to the  second order,

\begin{eqnarray}
\partial_\mu \phi_2
&=&\partial_\mu \left[\int_0^{s} ds_1 \mathscr{A}u(s_1) \phi_1(s_1)\right]\nonumber\\
&=&\int_0^{s} ds_1\partial_\mu \mathscr{A}(s_1)\phi_1(s_1)\nonumber\\
& &+\int_0^{s} ds_1 \mathscr{A}(s_1)\partial_\mu \phi_1(s_1)
\end{eqnarray}
Once more, using the differential chain type rule and the definition of  $F_{\mu\nu}$, we obtain
\begin{eqnarray}
& &\partial_\mu \phi_2\nonumber\\
&=&\int_0^{s} ds_1 \frac{d }{ds_1} \mathscr{A}_\mu(s_1)\phi_1(s_1)
+\int_0^{s} ds_1 \mathscr{A}(s_1)\partial_\mu \phi_1(s_1)\nonumber\\
& &+ \int_0^{s} ds_1 \left\{\frac{}{} \mathscr{F}_\mu(s_1)
+ig\left[ \mathscr{A}_\mu\left(s_1\right), \mathscr{A}\left(s_1\right)\right]\right\}\phi_1(s_1)
\nonumber\\
&=&\mathscr{A}_\mu\left(s\right) \phi_1(s)
-\int_0^{s} ds_1 \mathscr{A}_\mu\left(s_1\right)\frac{d \phi_1(s_1)}{ds_1}\nonumber\\
& &+\int_0^{s} ds_1 \mathscr{A}\left(s_1\right)\partial_\mu \phi_1(s_1)
\nonumber\\
& &+ \int_0^{s} ds_1 \left\{\frac{}{} \mathscr{F}_\mu(s_1)
+ig\left[ \mathscr{A}_\mu\left(s_1\right), \mathscr{A}\left(s_1\right)\right]\right\}\phi_1(s_1)\hspace{0.8cm}
\end{eqnarray}
Using the result of the first order (\ref{partial-1}) and the relation
\begin{eqnarray}
\frac{d \phi_1(s_1)}{ds_1}=\mathscr{A}(s_1)
\end{eqnarray}
we  have
\begin{eqnarray}
& &\partial_\mu \phi_2\nonumber\\
&=&-\int_0^{s} ds_1 \left[ \mathscr{A}_\mu\left(s_1\right),  \mathscr{A}\left(s_1\right)\right]\nonumber\\
& &+\mathscr{A}\left(s_1\right)\phi_1(s)-\phi_1(s) \mathscr{A}\left(0\right)\nonumber\\
& &+\int_0^{s} ds_1 \int_0^{s_1} ds_2 \mathscr{A}\left(s_1\right)\nonumber\\
& &\times\left\{\frac{}{} \mathscr{F}_\mu(s_2)
+ig\left[ \mathscr{A}_\mu\left(s_2\right), \mathscr{A}\left(s_2\right)\right]\right\}\nonumber\\
& &+\int_0^{s} ds_1 \int_0^{s_1} ds_2\nonumber\\
& &\times\left\{\frac{}{} \mathscr{F}_\mu(s_1)
+ig\left[ \mathscr{A}_\mu\left(s_1\right), \mathscr{A}\left(s_1\right)\right]\right\}\mathscr{A}\left(s_2\right)
\end{eqnarray}
Hence it is shown that we can obtain $\partial_\mu \phi_2$ by using the result of $\partial_\mu \phi_1$ in an iterative way.
Such process can be generalized to higher order case. For example,
\begin{eqnarray}
& &\partial_\mu \phi_{n+1}\nonumber\\
&=&\mathscr{A}_\mu\left(s\right) \phi_n(s)
-\int_0^{s} ds_1 \mathscr{A}_\mu\left(s_1\right)\frac{d \phi_n(s_1)}{ds_1}\nonumber\\
& &+\int_0^{s} ds_1 \mathscr{A}\left(s_1\right)\partial_\mu \phi_n(s_1)
\nonumber\\
& &+ \int_0^{s} ds_1 \left\{\frac{}{} \mathscr{F}_\mu(s_1)
+ig\left[ \mathscr{A}_\mu\left(s_1\right), \mathscr{A}\left(s_1\right)\right]\right\}\phi_n(s_1)\nonumber\\
\end{eqnarray}
Using  the general relation
\begin{eqnarray}
\frac{d \phi_{n+1}(s_{1})}{ds_1}=\mathscr{A}(s_1) \phi_{n}(s_{1})
\end{eqnarray}
yields
\begin{eqnarray}
& &\partial_\mu \phi_{n+1}\nonumber\\
&=&\mathscr{A}_\mu\left(s\right) \phi_n(s)-\int_0^{s} ds_1 \mathscr{A}_\mu\left(s_1\right) \mathscr{A}\left(s_1\right)\phi_{n-1}(s_1)\nonumber\\
& &+\int_0^{s} ds_1 \mathscr{A}\left(s_1\right)\partial_\mu \phi_n(s_1)
\nonumber\\
& &+ \int_0^{s} ds_1 \left\{\frac{}{} \mathscr{F}_\mu(s_1)
+ig\left[ \mathscr{A}_\mu\left(s_1\right), \mathscr{A}\left(s_1\right)\right]\right\}\phi_n(s_1).\nonumber\\
\end{eqnarray}
We can express $\partial_\mu \phi_n(s_1)$ in terms of the lower order terms $ \phi_{n-1}$ and $ \phi_{n-2}$.
\begin{eqnarray}
& &\partial_\mu \phi_{n+1}\nonumber\\
&=&\mathscr{A}_\mu\left(s\right) \phi_n(s)
+ \int_0^{s} ds_1 \mathscr{F}_\mu(s_1)\phi_n(s_1)\nonumber\\
& &+\int_0^{s} ds_1 \left[\mathscr{A}_\mu\left(s_1\right), \mathscr{A}\left(s_1\right)\right]\left(ig \phi_{n}(s_1)-\phi_{n-1}(s_1)\right)\nonumber\\
& &-\int_0^{s} ds_1\int_0^{s_1} ds_2 \mathscr{A}\left(s_1\right) \mathscr{A}_\mu\left(s_2\right) \mathscr{A}\left(s_2\right)\phi_{n-2}(s_1)
\nonumber\\
& &+ \int_0^{s} ds_1 \mathscr{A}\left(s_1\right) \left\{\frac{}{} \mathscr{F}_\mu(s_1)
+ig\left[ \mathscr{A}_\mu\left(s_1\right), \mathscr{A}\left(s_1\right)\right]\right\}\phi_{n-1}(s_1)\nonumber\\
& &+\int_0^{s} ds_1 \int_0^{s_1} ds_2 \mathscr{A}\left(s_1\right)\mathscr{A}\left(s_2\right) \partial_\mu \phi_{n-1}(s_2)
\end{eqnarray}
Continuing this iterative process we  finally have
\begin{eqnarray}
& &\partial_\mu \phi_{n+1}\nonumber\\
&=&\mathscr{A}_\mu\left(s\right)\phi_n(s)-\phi_n(s)\mathscr{A}_\mu\left(s\right)\nonumber\\
& &+ \int_0^{s} ds_1 \mathscr{F}_\mu\left(s_1\right)
+ \int_0^{s} ds_1 \int_0^{s_1} ds_2 \mathscr{A}\left(s_1\right)\mathscr{F}_\mu\left(s_2\right)
+\cdot\cdot \cdot\nonumber\\
& &+\int_0^{s} ds_1  \cdot \cdot \cdot \int_0^{s_n} ds_{n+1}
\mathscr{A}\left(s_1\right)\cdot \cdot \cdot \mathscr{A}\left(s_n\right)
\mathscr{F}_\mu\left(s_{n+1}\right)\nonumber\\
& &+\int_0^{s} ds_1 \left[ \mathscr{A}_{\mu}\left(s_1\right),  \mathscr{A}\left(s_1\right)\right]
\left(ig\phi_n(s_1)-\phi_{n-1}(s_1)\right)\nonumber\\
& &+\int_0^{s} ds_1 \int_0^{s_1} ds_2\mathscr{A}\left(s_1\right)\left[ \mathscr{A}_{\mu}\left(s_2\right),  \mathscr{A}\left(s_2\right)\right]\nonumber\\
& &\times\left(ig\phi_{n-1}(s_2)-\phi_{n-2}(s_2)\right)\nonumber\\
& &+\cdot \cdot\cdot \nonumber\\
& &+\int_0^{s} ds_1 \int_0^{s_1} ds_2\cdot \cdot\cdot \int_0^{s_n} ds_{n+1}\mathscr{A}\left(s_1\right)
\cdot \cdot \cdot \mathscr{A}\left(s_{n-1}\right)\nonumber\\
& &\left[ \mathscr{A}_{\mu}\left(s_n\right),  \mathscr{A}\left(s_n\right)\right]\left(ig\phi_{1}(s_n)-\phi_{0}(s_n)\right)\nonumber\\
& &+\int_0^{s} ds_1 \int_0^{s_1} ds_2\cdot \cdot\cdot \int_0^{s_n} ds_{n+1}\mathscr{A}\left(s_1\right)
\cdot \cdot \cdot \mathscr{A}\left(s_{n}\right)\nonumber\\
& &\times\left[ \mathscr{A}_{\mu}\left(s_{n+1}\right),  \mathscr{A}\left(s_{n+1}\right)\right]
ig \phi_{0}(s_{n+1})
\end{eqnarray}
Summing  all of them gives rise to the partial derivative of $\omega(s)$
\begin{eqnarray}
\partial_\mu \omega(s)=\sum_{n=0}^{\infty}\frac{(ig)^n}{n!} \partial_\mu \phi_n
\end{eqnarray}
It is found that all the commutation terms cancel each other and the final result reads
\begin{eqnarray}
\partial_\mu \omega(s)&=&\mathscr{A}_{\mu}\left(s\right) \omega(s) -\omega(s) \mathscr{A}_{\mu}\left(0\right)\nonumber\\
& &+\int_0^{s} ds_1 \omega^{-1}(s_1)\mathscr{F}_{\mu}\left(s_1\right)\omega(s_1)
\end{eqnarray}
Using the constraints (\ref{y0s}) and considering the assumption  $F^{\mu\nu}=0$, we  have
\begin{eqnarray}
\partial_\mu \omega=ig {A}_{\mu} \omega
\end{eqnarray}
Thus, we have verified that the expression (\ref{omega}) is the solution of differential equation (\ref{dif}) and the
integrability condition if $F_{\mu\nu}=0$. According to the theory of linear differential equation, this solution must be unique with some
specific initial condition which means that the solution (\ref{omega}) does not depend on the path we choose. This conclusion of path independence
can also be obtained from the non-Abelian Stokes Theorem \cite{Fishbane:1980eq,Hirayama:1999ar,Chen:2000aa}.

\section{summary}
\label{summary}
In the present work, we have reviewed some  issues which are very important when we deal with  the calculation in the light cone gauge.
First, we discussed why  we can not arbitrarily choose the boundary conditions of the gauge potential at the light cone infinity.
Then we showed  how  the singularities appear in the light cone gauge, how they are related to the boundary conditions.
and how we can regularize them in a proper  way corresponding to the different boundary conditions. Later on, we showed how to derive the
 gauge link at the light cone infinity from these singularities in Drell-Yan process. Finally we verified that  the gauge link at the light cone
 infinity  has no dependence on the path not only for the  Abelian field but also for non-Abelian gauge field.

\begin{acknowledgments}
 This work  was supported in part by the National Natural Science Foundation of China
under the Grant No.~11105137 and  CCNU-QLPL Innovation Fund (QLPL2011P01)
\end{acknowledgments}

\end{document}